\documentclass{PoS}
\usepackage{epsfig}
\usepackage{bm}
\usepackage{amssymb}
\usepackage{fontenc}
\usepackage{times}
\usepackage{mathptmx}
\usepackage{graphicx}

\title{Models of Walking Technicolor on the Lattice}

\ShortTitle{Models of Walking Technicolor on the Lattice}

\author{\speaker{D.~K.~Sinclair}%
         \thanks{This research was supported in part by US Department of Energy
         contract DE-AC02-06CH11357}
         \\
HEP Division, Argonne National Laboratory, 9700 South Cass Avenue, Argonne, 
Illinois 60439, USA\\
E-mail: \email{dks@hep.anl.gov}}

\author{J.~B.~Kogut\\
Department of Energy, Division of High Energy Physics, Washington, DC 20585,
USA\\
and\\
Department of Physics -- TQHN, University of Maryland, 82 Regents Drive, 
College Park, MD 20742, USA\\
        E-mail: \email{jbkogut@umd.edu}}

\abstract{We study QCD with 2 colour-sextet quarks as a walking-Technicolor
candidate. As such it provides a description of the Higgs sector of the standard
model, in which the Higgs field is replaced by the Goldstone `pions' of this
QCD-like theory, and the Higgs itself is the $\sigma$. Such a theory will need 
to be extended if it is to also give masses to the quarks and leptons. What we 
are attempting to determine is whether it is indeed QCD-like and hence walking, 
or if it has an infrared fixed point making it a conformal field theory. We do
this by simulating its lattice version at finite temperature and observing the
running of the bare (lattice) coupling at the chiral transition, as the lattice
spacing is varied, and comparing this running with that predicted by 2-loop
perturbation theory. Our results on lattices with temporal extents ($N_t$) up to
12 indicate that the coupling runs, but not as fast as asymptotic freedom 
predicts. We discuss our program for studying the zero-temperature 
phenomenology of this theory. 

\parindent 0.25in
QCD with 3 colour-sextet quarks, which is believed to be conformal is studied
for comparison. Simulations of this theory at finite temperature, on lattices
with $N_t$ as large as 8 indicate that the coupling still runs, and shows no
sign of approaching a finite limit for large $N_t$, in contrast to what is 
expected for a conformal theory. We are now extending these runs to $N_t=12$.
While it is too early to draw any conclusions from our $N_t=12$ runs, there is
a hint that the running of the coupling might have slowed.

Finally we looked at a new candidate theory, $SU(2)$ gauge theory with 3
Majorana adjoint fermions. However, this did not appear to allow embedding of
the electroweak group ($SU(2) \times U(1)$) to give the correct masses to the
$W$s and $Z$.
}

\FullConference{The 32nd International Symposium on Lattice Field Theory\\
                 23-28 June, 2014\\
                 Columbia University New York, NY}

\begin{document}

\section{Introduction}

We study models for the Higgs sector in which the Higgs is composite. In
particular, we study Technicolor models -- QCD-like theories with massless
fermions, where the Goldstone pion-like excitations play the role of the Higgs
field, giving mass to the $W^\pm$ and $Z$
\cite{Weinberg:1979bn,Susskind:1978ms}. Of particular interest are
walking-Technicolor models, where there is a range of mass scales over which
the running coupling evolves very slowly
\cite{Holdom:1981rm,Yamawaki:1985zg,Akiba:1985rr,Appelquist:1986an} Such
models can avoid the phenomenological problems with naive Technicolor.

QCD with 2 colour-sextet quarks is a candidate walking-Technicolor model. 
(For a summary of the properties of $SU(N)$ gauge theories which lead one to
such a conclusion see \cite{Sannino:2004qp,Dietrich:2005jn}.) We
need to distinguish whether this theory walks or is conformal. It is
attractive because it has just the right number of Goldstone bosons (3) to
give mass to the $W^\pm$ and $Z$, with none left over. Other groups are
studying this model, in particular, DeGrand, Shamir \& Svetitsky 
\cite{Shamir:2008pb,DeGrand:2008kx,DeGrand:2009hu,DeGrand:2010na,%
DeGrand:2012yq,DeGrand:2013uha}
and the Lattice Higgs Collaboration
\cite{Fodor:2009ar,Fodor:2011tw,Fodor:2012ty,Fodor:2012uw,Fodor:2014pqa,wong}
We study this theory at finite temperature to see if the coupling at the
chiral transition evolves as predicted by asymptotic freedom for a
finite-temperature transition \cite{Kogut:2011ty,Sinclair:2013era}.

QCD with 3 colour-sextet quarks, which is believed to be conformal, is studied
for comparison \cite{Kogut:2014kla}. For a conformal theory, the chiral
transition is a first-order bulk transition at fixed coupling separating the
chirally-symmetric conformal theory from the chirally broken phase, which is a
lattice artifact.

We simulate these theories, latticized with unimproved staggered fermions, using
the RHMC method.

If QCD with 2 sextet quarks walks, we need to answer the following questions.
Does QCD with 2 colour-sextet quarks have a light Higgs with standard-model
properties? What other light particles are in its spectrum? Can any of its
particles be dark-matter candidates? Do its $S$ (,$T$ and $U$) parameter(s)
pass precision electroweak tests.

We have also considered $SU(2)$ Yang-Mills with 3 Majorana/Weyl fermions.
However, we have been unable to embed the electroweak gauge group 
($SU(2) \times U(1)$) into this theory to give physical masses to the $W$s 
and $Z$.

\section{QCD with 2 colour-sextet quarks at finite temperature}

We simulate QCD with 2 color-sextet quarks at finite temperature by simulating
on an $N_s^3 \times N_t$ lattice with $N_s >> N_t$. Since $T=1/N_ta$,
increasing $N_t$ with $T$ fixed decreases $a$. Assuming the chiral phase
transition is a finite-temperature transition, this yields a convenient $T$,
$T_\chi$. Measuring $g$ or $\beta=6/g^2$ at $T_\chi$ gives a running coupling
at a sequence of $a$s which approach zero as $N_t \rightarrow \infty$.

\subsection{$N_t=12$}

Much of the past year has been devoted to increasing the statistics for our
simulations on $24^3 \times 12$ lattices at quark masses $m=0.0025$ and
$m=0.005$, close to the chiral transition. For our largest mass $m=0.01$
we have extended our simulations at low $\beta$s to determine the position
of the deconfinement transition.

Because the $\beta$ dependence of the chiral condensate is so smooth for
the masses we use, we determine the position $\beta_\chi$ of this
transition from the peaks in the (disconnected) chiral susceptibility:
\begin{equation}
\chi_{\bar{\psi}\psi} = V \left[\langle (\bar{\psi}\psi)^2 \rangle
                               -\langle \bar{\psi}\psi \rangle^2\right]
\end{equation}
extrapolated to $m=0$. $V$ is the space-time volume.

For $m=0.01$ in the range $5.7 \le \beta \le 5.9$, near the deconfinement
transition, we run for 50,000 trajectories for each $\beta$ with $\beta$s
spaced by $0.02$. In the range $6.6 \le \beta \le 6.9$, near the chiral
transition we run for 25,000 trajectories per $\beta$ with $\beta$s spaced by
$0.02$. Elsewhere in the range $5.7 \le \beta \le 7.2$ we run for 10,000
trajectories for $\beta$s spaced by $0.1$.

For $m=0.005$ in the range $6.6 < \beta \le 6.9$, we run for 50,000
trajectories per $\beta$ at $\beta$s spaced by $0.02$. At $\beta=6.6$ we run
for 100,000 trajectories. Elsewhere in the range $6.4 \le \beta \le 7.2$, we
run 10,000 trajectories per $\beta$ for $\beta$s spaced by $0.1$.

For $m=0.0025$ in the range $6.7 \le \beta \le 6.9$, we run for 100,000
trajectories per $\beta$ with $\beta$s spaced by $0.02$. In the range 
$6.6 \le \beta < 6.7$, we run for 50,000 trajectories per $\beta$. We are
currently extending our run at $\beta=6.68$ to 100,000 trajectories. Elsewhere
in the range $6.5 \le \beta \le 7.2$ we run 10,000 trajectories per
$\beta$ at $\beta$s spaced by $0.1$.

\begin{figure}[htb]
\parbox{2.9in}{
\epsfxsize=2.9in
\centerline{\epsffile{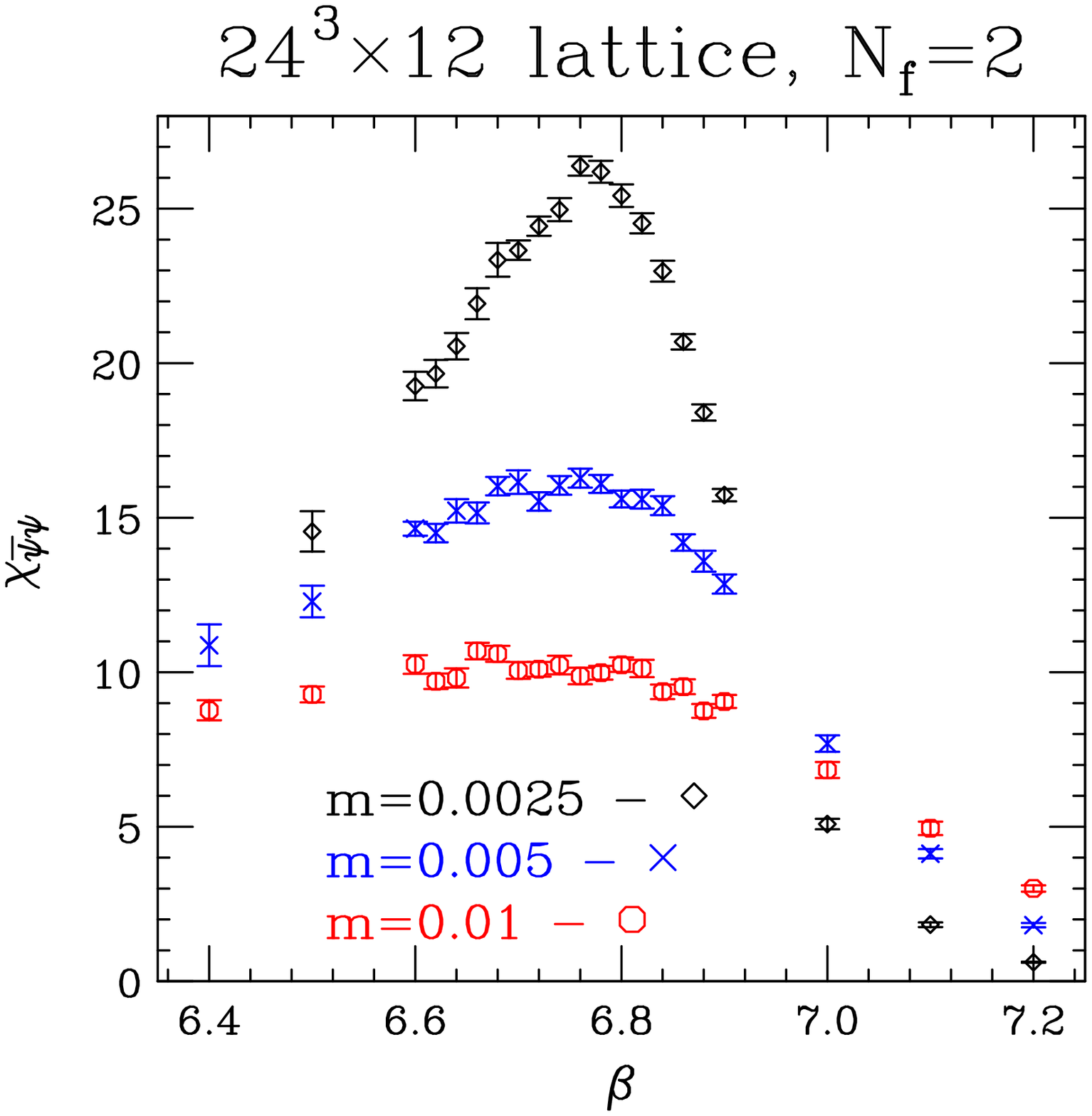}}
\caption{Chiral susceptibilities on a $24^3 \times 12$ lattice.}
\label{fig:chi12}
}
\parbox{0.2in}{}
\parbox{2.9in}{
\epsfxsize=2.9in 
\centerline{\epsffile{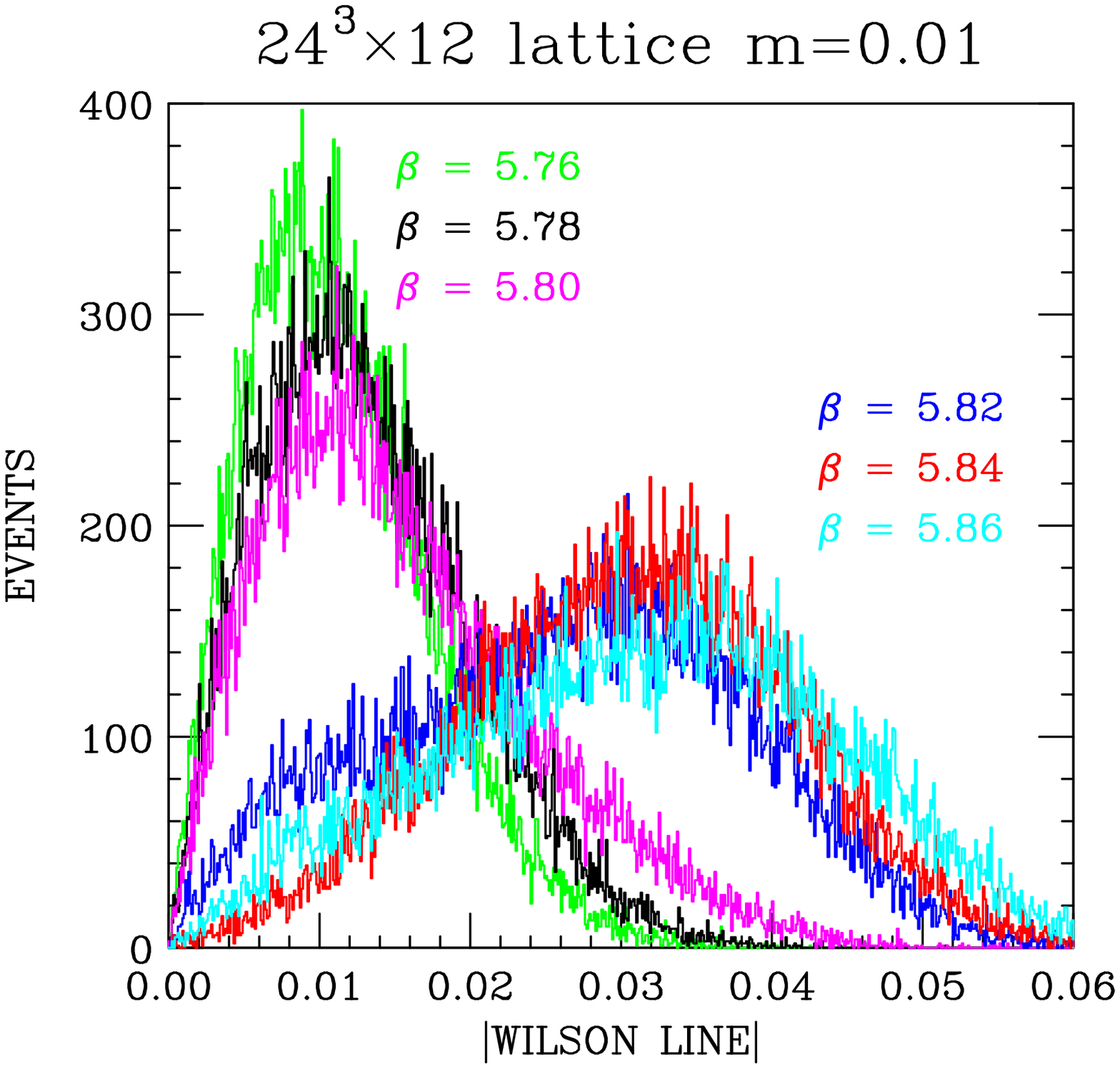}}
\caption{Histograms of magnitudes of Wilson Lines for $\beta$s close to
the deconfinement transition for $m=0.01$.}
\label{fig:wilhist} 
}
\end{figure}

While the chiral condensates measured in these simulations suggest that this
condensate will vanish in the chiral limit for large enough $\beta$ values,
they do not allow a precise determination of $\beta_\chi$ where this phase
transition occurs. For this we turn to the chiral susceptibilities.
Figure~\ref{fig:chi12} shows the chiral susceptibilities from these runs. The
peak of the $m=0.0025$ susceptibility yields an estimate of $\beta_\chi$,
namely $\beta_\chi=6.77(1)$. Combining this with our $N_t=8$ results
yields:
\begin{equation}
\beta_\chi(N_t=12)-\beta_\chi(N_t=8) = 0.08(2) \;,
\end{equation}
significantly smaller than the 2-loop perturbative prediction:
\begin{equation}
\beta_\chi(N_t=12)-\beta_\chi(N_t=8) \approx 0.12 \;.
\end{equation}

Figure~\ref{fig:wilhist} shows histograms of the magnitudes of Wilson Lines for
$m=0.01$ near to the deconfinement transition. From this we deduce that
$\beta_d = 5.81(1)$ for $m=0.01$. Previous experience indicates that this 
should be close to the value for $m=0$.

Table~\ref{tab:trans} shows our estimates for the positions of the deconfinement
($\beta_d$) and chiral ($\beta_\chi$) transitions from these and previous
simulations.
\begin{table}[h]
\centerline{
\begin{tabular}{|c|l|l|}
\hline
$N_t$   & \multicolumn{1}{c|}{$\beta_d$} & \multicolumn{1}{c|}{$\beta_\chi$} \\
\hline
4              &$\;$5.40(1)$\;$  &$\;$6.3(1)$\;$            \\
6              &$\;$5.54(1)$\;$  &$\;$6.60(2)$\;$           \\
8              &$\;$5.65(1)$\;$  &$\;$6.69(1)$\;$           \\
12             &$\;$5.81(1)$\;$  &$\;$6.77(1)$\;$           \\
\hline
\end{tabular}
}
\caption{$N_f=2$ deconfinement and chiral transitions for $N_t=4,6,8,12$.}
\label{tab:trans}
\end{table}

\section{Planned simulations and measurements}

We plan simulations at zero temperature to understand the phenomenology of this
theory, starting with simulations on a $36^3 \times 72$ lattice at $\beta$
values below the deconfinement transition for $N_t=36$. Unfortunately, such
$\beta$s are still too small to access the continuum limit. However, we hope
that we will be able to get results which are qualitatively correct. We then
plan to move to a $48^3 \times 96$ lattice which will allow us to move to
larger $\beta$s.

We will measure $f_\pi$ and the meson spectrum, including disconnected 
contributions. Here we will look for a light Higgs, and measure `taste'-breaking
in the `pion' spectrum. The `glueball' spectrum will be measured since we 
suspect that the low-lying glueballs could be light, and that mixing with the
mesons could produce a light Higgs-like particle. We will then need to check if
any Higgs candidate has the correct couplings to the $W^\pm$ and $Z$. We plan
to measure $S$-parameter contributions, and to determine the scaling behaviour 
of the chiral condensate to extract $\gamma_m$.

\section{QCD with 3 colour-sextet quarks at finite temperature}

We simulate lattice QCD with 3 colour-sextet quarks at finite temperature for
comparison with the 2-flavour case. This theory is believed to be conformal
with an infrared fixed point. The chiral transition should be a bulk
transition fixed at a finite constant $\beta_\chi$ for $N_t$ sufficiently
large. We have simulated this theory at $N_t=4$, $6$ and $8$, and are now
performing $N_t=12$ simulations.

For $N_t=6$ we simulate on a $12^3 \times 6$ lattice at $m=0.02$, $m=0.01$ and
$m=0.005$. Close to the chiral transition ($6.2 \le \beta \le 6.4$), at the
lowest quark mass ($m=0.005$), we simulate at $\beta$s separated by $0.02$,
with 100,000 trajectories per $\beta$. We estimate the position of the chiral
transition as the peak in the chiral susceptibility for $m=0.005$.

For $N_t=8$ we simulate on a $16^3 \times 8$ lattice at $m=0.01$ and
$m=0.005$. Close to the chiral transition ($6.28 \le \beta \le 6.5$) at the
lowest quark mass ($m=0.005$) we simulate at $\beta$s separated by $0.02$. with
100,000 trajectories per $\beta$. We estimate the position of the chiral
transition as the peak in the chiral susceptibility for $m=0.005$. The results
for the positions of the chiral and deconfinement transitions for $N_t=4$, $6$
and $8$ are given in table~\ref{tab:trans3}.
\begin{table}[h]
\centerline{
\begin{tabular}{|c|l|l|}
\hline
$N_t$  & \multicolumn{1}{c|}{$\beta_d$} 
       & \multicolumn{1}{c|}{$\beta_\chi$} \\
\hline
4              &$\;$5.275(10)$\;$  &$\;$6.0(1)  $\;$            \\
6              &$\;$5.375(10)$\;$  &$\;$6.278(2)$\;$            \\
8              &$\;$5.45(10) $\;$  &$\;$6.37(1) $\;$            \\
\hline
\end{tabular}
}
\caption{$N_f=3$ deconfinement and chiral transitions for $N_t=4,6,8$.
In each case we have attempted an extrapolation to the chiral limit.}
\label{tab:trans3}
\end{table}

\begin{figure}[htb]
\parbox{2.9in}{
\epsfxsize=2.9in
\centerline{\epsffile{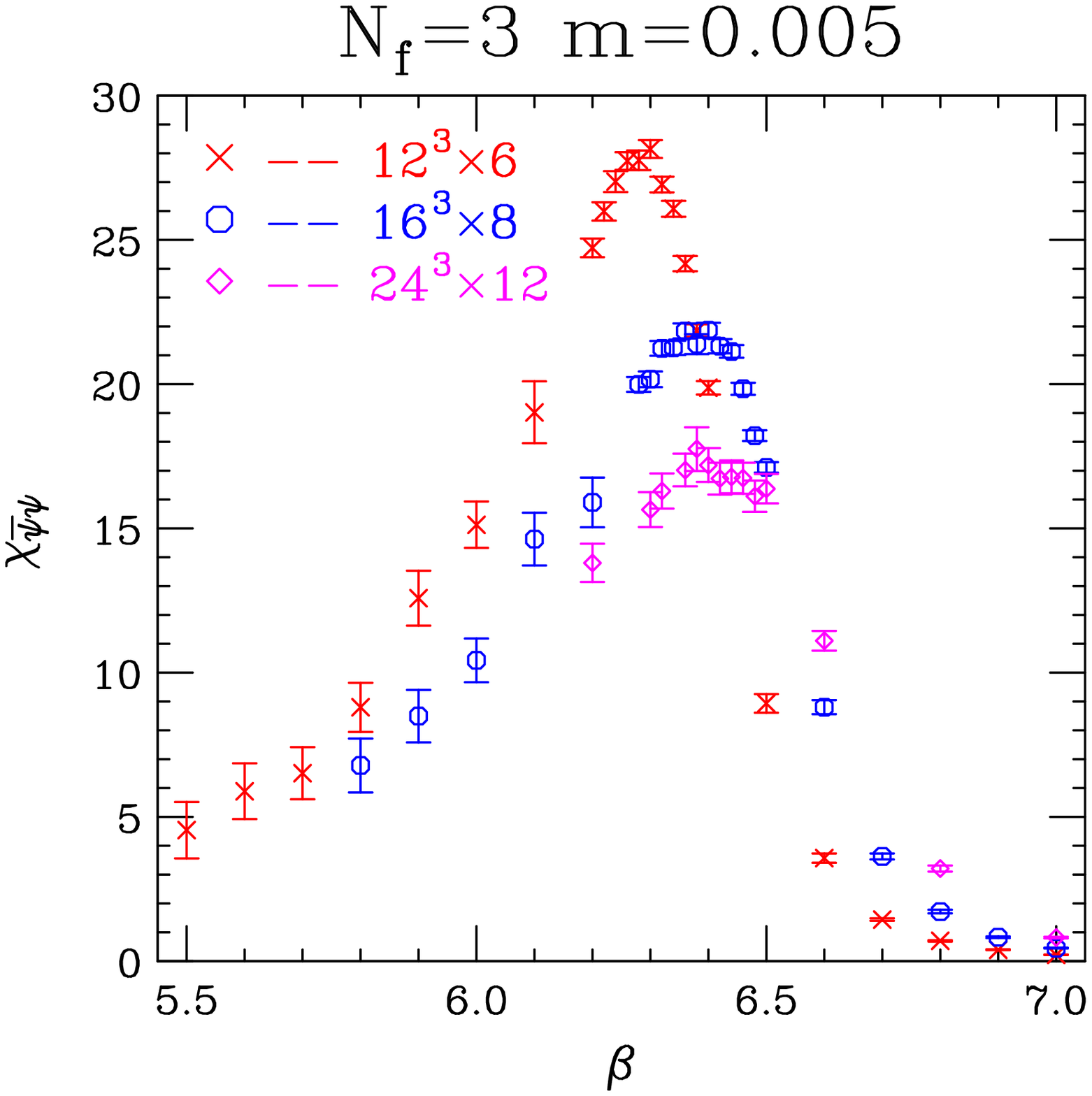}}
\caption{Chiral susceptibilities for $N_f=3$, $m=0.005$ on $12^3 \times 6$,
$16^3 \times 8$ and $24^3 \times 12$ lattices.}
\label{fig:chipbpm005}
}
\parbox{0.2in}{}
\parbox{2.9in}{
\epsfxsize=2.9in 
\centerline{\epsffile{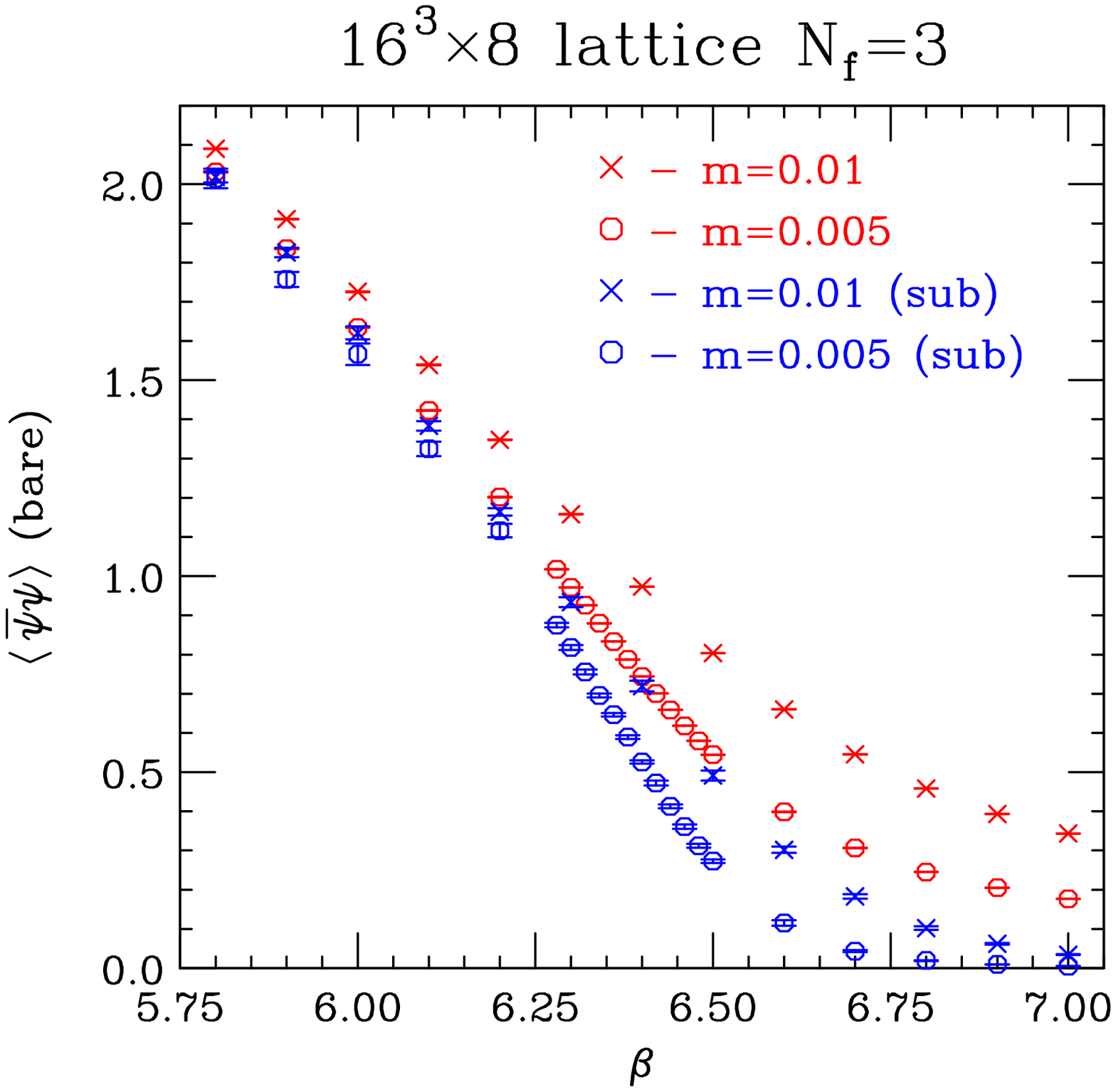}}
\caption{Chiral condensates on a $16^3 \times 8$ lattice for $m=0.005$ and
$m=0.01$. The red graphs are unsubtracted, lattice regulated condensates. The
blue graphs have been subtracted.}
\label{fig:pbp8}
}
\end{figure}

Since 
\begin{equation}
\beta_\chi(N_t=8) - \beta_\chi(N_t=6) = 0.09(1)
\end{equation}
we have yet to see evidence of a bulk transition. We are therefore starting
$N_t=12$ simulations on a $24^3 \times 12$ lattice.
Figure~\ref{fig:chipbpm005} shows the $m=0.005$ chiral susceptibilities
for $N_t=6$, $N_t=8$ and preliminary results for $N_t=12$.

Figure~\ref{fig:pbp8} shows the chiral condensates, both unsubtracted and
subtracted for our $16^3 \times 8$ simulations. The subtracted condensates
use the definition of the Lattice Higgs Collaboration \cite{Fodor:2011tw}:
\begin{equation}
\langle{\bar{\psi}\psi}\rangle_{sub} = \langle{\bar{\psi}\psi}\rangle 
-\left(m_V\frac{\partial}{\partial m_V}\langle{\bar{\psi}\psi}\rangle\right)
                                                                  _{m_V=m}\;,
\end{equation}
where $m_V$ is the valence-quark mass. Note, although it is clearer that the
subtracted condensate will vanish in the continuum limit for $\beta$
sufficiently large than is the case for the unsubtracted condensate, it still
does not yield an accurate estimate of $\beta_\chi$.

\section{$SU(2)$ gauge theory with 3 Majorana/Weyl colour-adjoint fermions}

The symmetries of this theory are easiest to see in terms of 2-component (Weyl)
fermions. 
\begin{equation}
{\cal L}=-\frac{1}{4}F^{\mu\nu}F_{\mu\nu}+\frac{1}{2}\psi^\dagger i \sigma^\mu 
\stackrel{\leftrightarrow}{D}_\mu \psi
+\frac{m}{2}\left[\psi^Ti\sigma_2\psi-\psi^\dagger i\sigma_2\psi^*\right] 
\end{equation}
where $\psi$ is a 3-vector in colour${_2}$ space and in flavour space.

If $m=0$, the chiral flavour symmetry is $SU(3)$. The Majorana mass
term reduces this flavour symmetry to the real elements of $SU(3)$, i.e. to
$SO(3)$. Thus when $m=0$ and the chiral symmetry breaks spontaneously, the 
chiral condensate is 
$\langle\psi^Ti\sigma_2\psi-\psi^\dagger i\sigma_2\psi^*\rangle$.
and the spontaneous symmetry breaking pattern is $SU(3) \rightarrow SO(3)$.
The unbroken generators of $SU(3)$ are the 3 imaginary generators. These form
a spin-1 representation under the unbroken $SO(3)$. The 5 broken generators
are the 5 real generators. They, as well as the 5 corresponding Goldstone
bosons, form a spin-2 representation of $SO(3)$.
 
The problem occurs when one tries to embed the weak $SU(2) \times U(1)$ group
in such a way as to give masses to $W^\pm$ and $Z$. This is easiest to see if
we consider the case where the Weinberg angle is zero. Then we need to embed
$SU(2)$ in such a way that all 3 components are broken spontaneously.
Thus we would need to make a set of $SU(2)$ generators from the 5 real $SU(3)$
generators. However, the $SU(2)$ algebra requires at least one of its
generators to be complex, so this is impossible. The only Weinberg angle which
would work is $\pi/2$ where the photon is pure $SU(2)$ and the $Z$ is pure
$U(1)$.

\section{Discussion and Conclusions}

We simulate lattice QCD with 2 colour-sextet quarks at finite temperature to
distinguish whether it is QCD-like and hence walks, or if it is a conformal
field theory. We run on lattices with $N_t=4,6,8,12$. $\beta_\chi$ increases by
$0.08(2)$ between $N_t=8$ and $N_t=12$. While this increase favours the
walking scenario, it is significantly smaller than the 2-loop prediction of
$\approx 0.12$. Is this because 2-loop perturbation theory is inadequate for
this lattice action and $\beta$? Are there sizable finite volume
corrections? Will the theory finally prove to be conformal?

If walking, this theory is a promising walking-Technicolor theory. It has 
just the right number of Goldstone bosons to give masses to the $W^\pm$ and $Z$,
with no extras to be explained away. We have outlined a program for checking
its zero-temperature properties. Does it have a light Higgs? Does it satisfy
the precision electroweak constraints? Does it have a Dark Matter candidate?
What about its particle spectrum?

We simulate QCD with 3 colour-sextet quarks which should be conformal. The
increase in $\beta_\chi$ between $N_t=6$ and $N_t=8$ is still appreciable
($0.09(1)$), so we don't yet have evidence for $\beta_\chi$ approaching a
finite constant as $N_t \rightarrow \infty$. We are now simulating at
$N_t=12$. This shows some promise.

QCD$_2$ with 3 Majorana/Weyl quarks does not appear to be a Technicolor
candidate.

\section*{Acknowledgements}
These simulations were performed on Hopper, Edison and Carver at NERSC, and
Kraken at NICS and Stampede at TACC under XSEDE project TG-MCA99S015,
and Fusion and Blues at LCRC, Argonne. NERSC is supported by DOE contract
DE-AC02-05CH11231. XSEDE is supported by NSF grant ACI-1053575.

\end{document}